\documentclass[twoside,article,norunningheads,natbib,a4paper]{./styles/svproc}

\usepackage{cancel}
\usepackage[normalem]{ulem}
\usepackage{algorithm}
\usepackage{algorithmicx}
\usepackage{algpseudocode} 
\usepackage[cmex10]{amsmath}
\usepackage{mathtools}
\usepackage{amssymb}
\usepackage{balance}
\usepackage{eqparbox}
\usepackage{bbm, flushend}
\usepackage{mathtools}
\usepackage[colorinlistoftodos]{todonotes}
\usepackage{color}
\usepackage{subcaption}
\usepackage{float}
\usepackage{setspace}
\usepackage{listings}

\newcommand{\bseq}{\begin{subequations}}
\newcommand{\eseq}{\end{subequations}}
\newcommand{\baln}{\begin{align}}
\newcommand{\ealn}{\end{align}}
\newcommand{\balnd}{\begin{aligned}}
\newcommand{\ealnd}{\end{aligned}}
\newcommand{\beq}{\begin{equation}}
\newcommand{\eeq}{\end{equation}}
\newcommand{\beqn}{\begin{eqnarray}}
\newcommand{\eeqn}{\end{eqnarray}}
\newcommand{\beqno}{\begin{eqnarray*}}
\newcommand{\eeqno}{\end{eqnarray*}}
\newcommand{\bma}{\begin{displaymath}}
\newcommand{\ema}{\end{displaymath}}
\newcommand{\bnu}{\begin{enumerate}}
\newcommand{\enu}{\end{enumerate}}
\newcommand{\bce}{\begin{center}}
\newcommand{\ece}{\end{center}}
\newcommand{\btb}{\begin{tabular}}
\newcommand{\etb}{\end{tabular}}
\newcommand{\ba}{\begin{array}}
\newcommand{\ea}{\end{array}}

\newtheorem{assumption}{Assumption}
\usepackage[numbers]{natbib}
\usepackage{url}

\begin{document}
\mainmatter              
\title{Prediction based computation offloading and resource allocation for multi-access ISAC enabled IoT system}
\titlerunning{Prediction based resouce allocation}  
%
\author{Duc-Thuan Le}
\authorrunning{Duc-Thuan Le} 
%
\tocauthor{Duc-Thuan Le}
\institute{HCMC University of Technology, VNU-HCM, Vietnam\\
}

\maketitle              

\begin{abstract}
In the new era of the Internet of Things (IoT), tasks are now being migrated to edge sites closer to data generators.
Mobile devices inherently encounter limitations in terms of energy and computational processing capabilities. 
In high mobility paradigm, ISAC provides a promising foundation for integrating deployment management within dynamic spatial settings. 
We are interested in applying prediction mechanism to resource allocation management by extracting data attributes, focusing on ISAC related contexts of the trajectory and velocity and making the allocating decision.  
We present a system design, a theoretical framework and an implementation of the ClusterMan software package. The numerical suggest that the strong clustering subset of feature may yield high accuracy up to 97\% in the prediction results.
\end{abstract}
\linespread{1.4}
\section{Introduction}
With the advent of the Internet of Thing, there has been a paradigm shift towards relocating tasks from centralized processing centers to the edge locations, where data is generated, has occurred and being exploded. This potential unlocks the high dynamic scenarios, such as automotive and industrial applications, which have previously been limited due to the problem of battery capacity and slow processor speed. When combined with spatial factors, as seen in Integrated sensing and communication (ISAC) \cite{xiong2023fundamental}, it forms a cornerstone in ongoing 6G implementations. However, employing an efficient resource allocation mechanism  in combination with understanding the spectral efficiency in spatial variance in the wireless channel remains incomplete.

The issue of offloading computation to servers has been researched and implemented across various system types \cite{kovachev2012adaptive}. In distributed and multi-access networks, objects can also be moved at the environmental level, such as with Docker \cite{ma2017efficient}. However, these mechanisms are constrained by the large volume of data, which is not always suitable for IoT system environments.
An alternative direction being considered is an adaptation of previous telecommunications device subsets, now applied to IoT, known as Mobile Edge computation offloading \cite{mach2017mobile}. In this approach, each data bit is carefully weighed against the energy consumed for processing at the device or at the server system \cite{nguyen2019computation}. While multi-site cloud computing \cite{nguyen2020joint} offers significant computational power, the data transmission latency is too large to operate matching the requirement of processing time delay of IoT devices. Therefore, mobile networks are utilized to supplement computational capabilities.

Current solutions in computation-assisted mobile network encounter the problem of spectral efficiency in spatial variances especially in high mobility paradigm. 
Consequently, the development of 6G networks aims to integrate sensing into communications and target ISAC enabled IoT system. 
The biggest challenge of handling spatial factors and trajectory movements to modulation speed is addressed in OTFS \cite{hadani2017orthogonal}. 
OTFS is noted for its ability to transmit data for shorter period of times, enabling longer-range radar and/or faster target tracking rates \cite{raviteja2019orthogonal}, making it a promising mechanism for ISAC.
We consider the offloading in the  practical implementation of ISAC, OTFS, specifically using velocity modulation, which poses a challenge due to the complexity of the modulation domain. Previous solutions focused on optimization approach, but failed to ensure the flexibility of the solution, depending on the nature of the data. 

This study applies attribute analysis and data correlation in building predictive solutions in Meta-learning (MAML) \cite{rajeswaran2019meta} to the problem of strategy selection and resource allocation mechanism construction. We present the two models relevant to the ISAC enable communication in IoT system
In the first model, we analyse the data structure and clustering characteristics. In the second model, we employ the regression model to formulate the clustered feature into resource allocation prediction. We examine different subset of resources includes frequency, speed and offloading factor. We developed a novel outlier theoretical framework to perform the prediction in spectral efficiency  and ensure the convergence of the proposed mechanism. We prove and show by numerical simulation that the capability prediction accuracy.

\section{System design}
\vspace{-0.4 cm}\subsection{ClusterMan architecture design}\vspace{-0.21 cm}
\label{sec:sysArch}
The architecture of the ClusterMan system is designed for deployment in a distributed system of mobile devices that require low-latency responsiveness and energy-efficient consumption. It employs wireless communication channels that leverage velocity modulation and a unified centralized processing server system. From a user's perspective, this system appears as a single server with unlimited capacity. Specifically, the described architecture in Figure \ref{fig:SysArchDiag} includes the following components:

\begin{figure}[h]
\centering
\vspace{0.2 cm}
\includegraphics[width=0.9\textwidth]{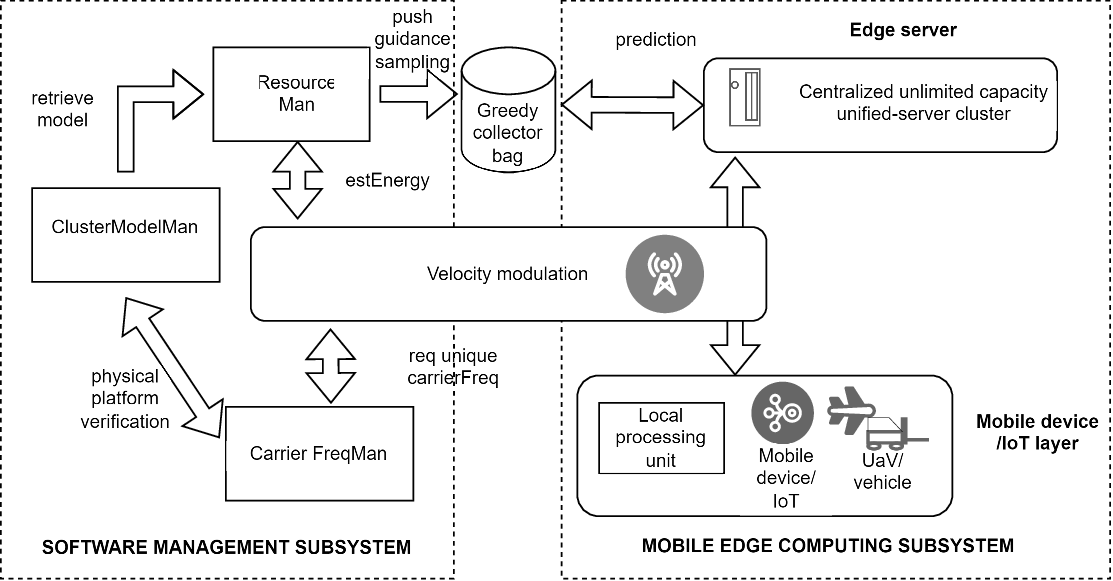}
\vspace{-0.2 cm}
\caption{System Architecture Diagram} 
\vspace{-0.2cm}
\label{fig:SysArchDiag}
\vspace{-0.5 cm}
\end{figure}

\textbf{Software Management Subsystem} orchestrates the efficient utilization of resources, manages cluster models, and optimizes carrier frequencies. It consists of three key components: the ResourceMan, the ClusterModelMan, and the CarrierFreqMan, detailed in the following list:

\vspace{-0.2 cm}
\begin{itemize}
    \item[-] \textbf{ResourceMan}: This component provides the allocation and utilization of available resources within the system. It monitors various system resources such as CPU, memory, storage, and network bandwidth to ensures that resources are efficiently allocated to different tasks within the system.
    \item[-] \textbf{ClusterModelMan}: This component manages and optimizes cluster configurations. It maintains a repository of diverse cluster models, and facilitates the selection of an appropriate model based on considerations such as performance goals, resource constraints, and scalability requirements.
    \item[-] \textbf{CarrierFreqMan}: This component focuses on managing carrier frequencies within wireless communication systems, including tasks such as frequency allocation, channel selection, and velocity modulation. It collects and analyzes wireless physical resource, then it determines the optimal carrier assignments to maximize resource utilization.
\end{itemize}

\vspace{-0.2 cm}
\noindent
\textbf{Mobile Edge Computing Subsystem} collaborates the two main components to enable efficient task processing and offloading, leveraging both centralized and local computing resources. The main components are:

\vspace{-0.2 cm}
\begin{itemize}
    \item[-] \textbf{Edge Server}: This component serves as a centralized computing node, equipped with unlimited processing capabilities to manage intensive computation tasks. It offers high-performance computing resources capable of handling multiple concurrent requests from mobile devices.
    \item[-] \textbf{Mobie Device/IoT layer}: This component represents the end-user device, such as mobile devices, vehicles, or IoT devices equipped with sensors, user applications enabling them to collect local data.
    These devices typically have limited computational resources and power constraints compared to Edge Servers then 
    thus requiring them to offload their heavy tasks to the Edge Server for further processing.
\end{itemize}

\vspace{-0.2 cm}
\noindent
\textbf{Velocity modulation library implementation}
employs Zak-based transformation in \cite{zak_transform} and provides the API $calcSE(\cdot)$ as described in the Listing \ref{lst:calSElib}.
\begin{lstlisting}[basicstyle=\linespread{0.9}\small,language=Python,label={lst:calSElib}, caption=velocity modulation library, frame=single,escapechar=\%]
import random
import math
import matplotlib.pyplot as plt
import numpy as np

# Caching Spectral Efficiency
SE = {}
bandwidth = 1E6
num_usr = 5
time_frm = 10E-3   # 10 ms
delta_f = 100E3      # 100 kHz
lightSpeed = 3E8

def calc_SE(speed, carrierFreq):
    global SE,bandwidth,num_usr,time_frm,delta_f,lightSpeed
    # Return cached version
    if speed in SE:
        return SE[speed]
    # Construct representation based on Zak-transform as in %\cite{zak_transform}%.
    ...
end
\end{lstlisting}
\vspace{-0.4 cm}
\subsection{ClusterMan interface and interaction specification}\vspace{-0.21 cm}
Within the Software Management Subsystem, the interaction among its components facilitates effective operation, efficient resource utilization, enhanced system performance, and reliable communication, includes:
\vspace{-0.2 cm}
\begin{itemize}
    \item[-] \textbf{Model retrieving}: the ResourceMan interacts with the ClusterModelMan to retrieve information about available cluster configurations and their associated features. This interaction enables the ResourceMan to make informed decisions regarding resource allocation and scheduling.
    \item[-] \textbf{Physical platform information retrieving} the ClusterModelMan, upon receiving a request from the ResourceMan, selects an appropriate cluster model based on the workload characteristics and resource requirements. It further leverages information from the CarrierFreqMan to establish the appropriated clustering scheme on the physical platform.
    \item[-] \textbf{Physical platform information provider} the CarrierFreqMan provides essential information to ClusterModelMan about the physical platform, including channel conditions and interference levels. This information is utilized to configure cluster settings.
\end{itemize}
\vspace{-0.2 cm}

\noindent
The interaction between the Edge Server and the Mobie Device/IoT layer is described as follows:

\vspace{-0.1 cm}
\begin{itemize}
    \item \textbf{Task offloading}: when the Mobie Device/IoT layer has tasks that exceed its processing capacity or require centralized resources, it delegate these tasks to the Edge Server. 
    This offloading process can harness the advantages of event communication systems in tandem with application migration \cite{nguyen2012ebc,chen2019dynamic} to orchestrate task and delegate tasks efficiently.
    \item \textbf{Centralized cluster processing}: the Edge server receives task requests from Mobile Devices and performs centralized cluster processing based on predictive estimation to make the resource allocation decision. Through resource orchestration, it leverages a unified unlimited computational resources at large scale \cite{sonkoly2020scalable} to execute heavy computations efficiently in almost zero-delay response.
    \item \textbf{Result delivery} when the Edge Server completes processing tasks, it sends the results back to the requesting Mobile Devices. These outcomes may include aggregated info and usually has small data size and can be omitted in order to decouple the transmission process and computation process. \cite{suganuma2018multiagent}.
\end{itemize}

\vspace{-0.5 cm}
\section{System Model}
\vspace{-0.4 cm}\subsection{System notations}\vspace{-0.21 cm}

Given a system consisting a set of $N$ users and one Edge Server (ES). Each user uses exactly one mobile device (MD) then there are total of $N$ MD in the system. On each device, there exists a set of computational tasks. 
These tasks process an input stream of data, and the total number of tasks is denoted by $K$, where each task is indexed by $k \in \{1,2,\dots,K\}$. As a result, input data of a $k$-th task on MD $n$ has the length of $D_{n,k}$ bits.

\vspace{-0.4 cm}\subsection{Task model}\vspace{-0.21 cm}

\noindent
Each input data of a task on a MD \( n \in N \) has a task with an offloading ratio \( l_{n,k} \in [0,1] \). That portion \( l_{n,k}D_{n,k} \) is sent to the ES for computation, thus not using the MD's resources. Consequently, the length of the data kept for local execution on the MD is \( (1 - l_{n,k})D_{n,k} \).

We ignore the time and energy consumption in the ES when processing the data \(l_{n,k}D_{n,k}\) of the task when offloading it to the ES for processing, because the processing capacity of the ES is infinite, so the execution time is fast and the resource usage is negligible and can be ignored.

\vspace{-0.4 cm}\subsection{Local computation model}\vspace{-0.21 cm}
\label{sec:LocalComp_model}
Each MD is assumed to operate at a fixed processing speed of \(f_n\) during the execution of any task.
To perform computation, the quantity describing the complexity is \(c_{n,k}\) with units of CPU cycles/bit, indicating the average number of clock cycles required to process one bit of input data.
The execution time of the task computed locally on the MD, \( T_{local} \), in seconds, is expressed as follows:

\vspace{-0,3 cm}
\begin{equation}
    \label{eqn:t_local}
    T_{local(n,k)} = \dfrac{c_{n,k} (1 - l_{n,k}) D_{n,k}}{f_{n}}
\vspace{-0.2 cm}
\end{equation}
\noindent
where \(f_{n}\) is the CPU speed of MD \(n\) measured in (Hz), and \(c_{n,k}\) is the number of CPU cycles required to execute one bit of data on MD \(n\) for the task $k$. 

The corresponding amount of energy is:

\vspace{-0,3 cm}
\begin{equation}
    \label{eqn:e_local}
    E_{local(n,k)} = \epsilon_{n} c_{n,k} f_{n}^2 (1 - l_{n,k}) D_{n,k}
\end{equation}
\noindent
where $\epsilon_{n,k}$ is a energy coefficient determined by the chip architecture of MD $n$.

\vspace{-0.4 cm}\subsection{Communication model}\vspace{-0.21 cm}
\label{sec:offloading_model}

Considering that the data size of the task processed result is negligible compared to the data size of the tasks themselves, in this work, we only focus on the uplink transmission. We consider the situation where the ES allocate spectrum bandwidth $W_n$ represents the bandwidth allocated to MD \(n\) by the ES to offload its computation. The uplink transmission rate from MD \(n\) to the edge ES can be rewritten as follows:
\vspace{-0.5 cm}
\begin{equation}
    \label{eqn:comm_rate}
    R_n = W_n \log_2 \left( 1 + \dfrac{p_n h_n}{\sigma^2} \right) = W_n {calSE}(v_k, f_c)
\vspace{-0.2 cm}
\end{equation}
\noindent
where \(p_n\) is the uplink transmission power of MD \(n\). Assume the paths between transmitter and receiver has little fading effect, the channel gain \(h_n\) is a constant close to 1. \(\sigma^2\) refers to the variance of the additive white Gaussian noise. The notation \(C_{{Zak}} = {calSE}(v_n, f_c)\) (bit/s/Hz) is the spectral efficiency (SE) with variables \(v_n\) being the vehicle speed and \(f_c\) being the carrier frequency.

We assume that the wireless links transmit data at a fixed rate measured in bits per second (bps). The transmission delay from MD \(n\) to the ES for offloading tasks can be defined as

\vspace{-0.5 cm}
\begin{equation}
    \label{eqn:t_trans}
    T_{off(n,k)} = \dfrac{l_{n,k} D_{n,k}}{R_{n}} = \dfrac{l_{n,k} D_{n,k}}{W_n {calSE}(v_k, f_c)} 
\vspace{-0.2 cm}
\end{equation}

\noindent
The corresponding energy to perform the transmission is:

\vspace{-0,5 cm}
\begin{equation}
    \label{eqn:e_trans}
    E_{off(n,k)} = p_n T_{off(n,k)} = \left( 2^{{calSE}(v_k, f_c)} - 1\right) \dfrac{\sigma^2}{h_n} \dfrac{l_{n,k} D_{n,k}}{W_n {calSE}(v_k, f_c)}
\vspace{-0.2 cm}
\end{equation}

\vspace{-0.4 cm}\subsection{Problem modeling statement}\vspace{-0.21 cm}
\label{sec:DesignModelProblemStatement}
The estimated total execution time for processing the task $k$ of MD \(n\) for any case can be expressed as follows:

\vspace{-0,8 cm}
\begin{equation}
    \label{eqn:total_time}
    T_{n,k} = T_{off(n,k)} + T_{local(n,k)} =\dfrac{l_{n,k} D_{n,k}}{W_n {calSE}(v_k, f_c)}  + \dfrac{c_{n,k} (1 - l_{n,k}) D_{n,k}}{f_{n}}
\vspace{-0,3 cm}
\end{equation}

The total energy consumption for processing the task of MD \(n\) at time \(k\) is the sum of the component energies
\vspace{-0,3 cm}
\begin{equation} 
\label{formula:e_total}
E_{n,k} = E_{off(n,k)} + E_{local(n,k)}
\vspace{-0,3 cm}
\end{equation}

By substituting the computed quantities from expressions (\ref{eqn:e_local}) and (\ref{eqn:e_trans}) to the equation (\ref{formula:e_total}), we obtain the formulation of the total energy expression optimization problem as follows:

\textbf{(Problem P1):}\\
\vspace{-0.7 cm}
\begin{align*}
    \begin{split}
        &\min\limits_{\Omega_1} \sum_{n\in \mathcal{N}} \sum_{n\in \mathcal{K}} \left( 2^{{calSE}(v_k, f_c)} - 1\right) \dfrac{\sigma^2}{h_n} \dfrac{l_{n,k} D_{n,k}}{W_n {calSE}(v_k, f_c)} + \epsilon_{n} c_{n,k} f_{n}^2 (1 - l_{n,k}) D_{n,k}
    \end{split}
\end{align*}

\vspace{-0.2 cm}
where the set of optimization variables is defined as $\Omega_1{=}\{f_n,l_{n,k},v_n,f_c\}$ and the computation of \({calSE}(\cdot)\) is considered an indivisible unit task.
In the subsequent steps, we will develop the solution for problem \textbf{(P1)}.

\section{Proposed Algorithm}
\vspace{-0.3cm}
\label{sec:DesignAlgEnhClusPre}
We aim to develop a prototype system where the components of the system are modeled and tested, then a feasible solution algorithm is provided. The algorithms designed at this stage have the nature of finding solution approaches to demonstrate the feasibility of the topic and is far from optimum. The results of the algorithm, named \textbf{IteraGAlg}, are described in section \ref{sec:DesignAlgGreedy}, it consists of two steps: configuration setup and greedy solution selection. 

In the next stage, the enhanced algorithm design are shaped by the analysis of the feature selection. 
Despite providing a solution with low complexity, the greedy yields results that diverge significantly from the actual system configuration. In some cases, the difficulty of the problem formulation and the high dimensions of problem spaces make it difficult for us to find the optimum. 
In our extended approach, we employ predictive methods to augment the baseline solution by leveraging the adaptation from an unrealistic high-dimensional problem to a realistic data-driven approach.
Therefore,
we proposed an algorithm in section \ref{sec:DesignClusPreAlg}, named \textbf{ClusPreAlg}, that \textit{does not aim to} find \textit{an optimum}, instead we apply the prediction to provide adaption to the \textit{realistic experimental environment} with large scalability and heterogeneity. 

\subsection{The Greedy Algorithm - IterGAlg}
\label{sec:DesignAlgGreedy}
From the formulation of \textbf{Problem (P1)} includes task model, local execution model, and offloading model, we propose a greedy algorithm that optimizes the total energy consumption of all MD and tasks. Assuming a pool of MD and tasks of individual MD is already given as well. The algorithm details are presented in Algorithm \ref{alg:greedyOffload} is processed through the 3 steps as listed below:

\vspace{-0.4cm}
\subsubsection{Step FI - Feasible Initialization} 
In this step, we initialize the system with a feasible set of values. As is common in offloading problems, we are interested in starting the offloading at the 50\% of the data input size. This initialized value might be changed depending on the different test scenarios later.

\vspace{-0.4cm}
\subsubsection{Step FS - Feasible Solution} 
This step performs the initialized calculation of total energy with a predefined offload list as an input argument. There are two steps of calculations, which include the local execution energy and the offloading energy energy. The former term is straightforward, achieved by applying mathematical modeling, while the latter term involves the formulation of offloading transmission and employs the external determination of spectral efficiency, as mentioned in \ref{sec:offloading_model}.

\vspace{-0.4cm}
\subsubsection{Step GIm - Greedy Improvement} 
The greedy improvement step starts by entering a loop where it tries to improve the result using greedy approaches after each iteration. We keep track of the best reachable result in a global optimum value. Then, in each loop step, we recalculate the new total energy consumption of all tasks. There are two expected outputs:

\textbf{An improvement is obtained}
if a updated total energy results is less than the current tracking minimum, the algorithm updates this minimum value. In addition to updating the new minimum energy consumption, the algorithm picks out the worst-performed task, i.e. the task consuming most energy, and increases it offloading ratio by 0.01 or 1\% to seek for energy consumption reduction and continues the next iteration.

\textbf{A saturated stage is reach; no improvement is found}
if this stage is reached, it indicates that the algorithm has exhausted all feasible options for improvement without achieving any further enhancements. Due to the step searching approaches, the algorithm has performed a semi-brute-force iterations to explore all the possibilities and surmount the limitations inherent to the current approach.

\noindent
\vspace{-0.8cm}
\begin{algorithm}[H]
    \caption{Algorithm \textbf{IteraGAlg} based on Iterative greedy decision} 
    \label{alg:greedyOffload}
    \small 
    \begin{algorithmic}[1]
        \State \textbf{Initialize:}
        \State \quad $\textit{cycle\_per\_bit} \gets \text{array} [1..K]$ \quad \quad \quad \# cycle per bit for all tasks
        \State \quad $\textit{coefficient} \gets \text{array} [1..K]$ \quad \quad \quad \# MD energy coefficient for all tasks
        \State \quad $\textit{frequency} \gets \text{array} [1..K]$ \quad \quad \quad \# MD processing frequency of all tasks
        \State
        \State \quad $\textit{se\_list} \gets \text{array} [1..K]$ \quad \quad \quad \# spectral efficiency for all tasks
        \State \quad $\textit{bandwidth\_list} \gets \text{array} [1..K]$ \quad \quad \quad \# communication bandwidth for all tasks
        \State \quad $\textit{gain\_list} \gets \text{array} [1..K]$ \quad \quad \quad \# communication gain for all tasks
        \State \quad $\textit{noise\_power} \gets \text{array} [1..K]$ \quad \quad \quad \# communication noise power for all tasks
        \State
        \State \quad $\textit{offload\_list} \gets \text{array} [1..K]$ \quad \quad \quad \# offloading ratio for all tasks
        \State \quad $\textit{data\_size} \gets \text{array} [1..K]$ \quad \quad \quad \# data size for all tasks
        \State
        \State \quad $\textit{local\_energy} \gets \text{array} [1..K]$ \quad \quad \quad \# local energy for all tasks
        \State \quad $\textit{offload\_energy} \gets \text{array} [1..K]$ \quad \quad \quad \# offload energy for all tasks
        \State
        \Function{get\_total\_energy}{offload\_list}
            \State $\textit{local\_size} {=} (1 {-} \textit{offload\_list}) {*} \textit{data\_size}$

            \State $\textit{offload\_size} {=} \textit{data\_size} {-} \textit{local\_size}$
        
            \State $\textit{local\_energy} {=} \textit{local\_size}{*} \textit{cycle\_per\_bit} {*} \textit{coefficient} {*} \textit{frequency} {*} \textit{frequency} $
            
            \State $\textit{offload\_energy} {=} (2^{\text{SE}} {-} 1) {/} \textit{SE} {*} (\textit{noise} / (\textit{gain} {*} \textit{bandwidth})) {*} \textit{offload\_size}$
        
            \State \Return $\textit{local\_energy} + \textit{offload\_energy}$
        \EndFunction

        \Function{optimize}{}
            \State \textbf{Step FI:}\;\;\;$\textit{offload\_list} = [0.5] * N; \textit{best\_so\_far} = +inf$
            \State \textbf{Step FS:}\;\;\;$\textit{total\_energy} = GET\_TOTAL\_ENERGY(offload\_list)$
            
            \State \textbf{Step GIm:} 
            \While{(SUM(\textit{total\_energy}) < \textit{best\_so\_far})}
                \State \textit{best\_so\_far} = SUM(\textit{total\_list})
                \State \textit{offload\_list}[INDEX(MAX(total\_list))] += 0.01
                \State \textit{total\_energy} = GET\_TOTAL\_ENERGY(offload\_list)
            \EndWhile
        \EndFunction
    \end{algorithmic} 
\end{algorithm}

\subsection{Clustering Prediction Approach Algorithm - ClusPreAlg}
\label{sec:DesignClusPreAlg}
The greedy technique often produces output that is significantly different from the actual system. There are several reasons for this disparity, such as the complexity of problem formulation model and the large size of problem domains. In certain scenarios, navigating through input variants poses significant challenges in achieving the optimal solution.
We utilize two transformed phases to improve the baseline solution, thereby enhancing its adaptivity.

\vspace{-0.3cm}
\begin{itemize}
\item \textbf{Feature extracting} 
we tackle the disparity by a refined methodologies to extract the feature classification which is capable of better approximating real-world scenarios

\item \textbf{Clustering prediction} after obtaining the extraction result, we aim to enhance the baseline solution by employing clustering prediction techniques. The prediction with pragmatic data-driven approach holds the potential to bridge the gap between theoretical models and real-world applications.
\end{itemize}

The whole process can be summarized in the pipeline workflow described in Section \ref{sec:pipelineworkflow}. Details of the first transformed phase are provided in Section \ref{sec:DesignAlgClusterPhase} and are mainly focus in the function $TrainClusteredModels(\;\;)$. The description of the second transformed phase, which involves mitigating feature extraction to prediction, is implemented in the function $ClusteringPrediction(\;\;)$ and is presented in Section \ref{sec:DesignAlgClusterPredict}. To provide the overview picture, we describe the procedure in Algorithm \ref{alg:cluster_predict}. To enhance theoretical contribution, we draw the conclusion in the Remark \ref{theory:remarkStrongClusterImpact} and then make a statement of the Proposition \ref{theory:ClusteringProposition}.

\vspace{-0.2 cm}\subsubsection{The prediction and pipeline approach}\vspace{-0.21 cm}
\label{sec:pipelineworkflow}

The proposed solution starts by simulating a set of random MD and tasks and aims to optimize the total energy consumption among all tasks of devices. Next, the optimized tasks as well as MD information are compiled into a dataset.

A feature selection process is applied on the dataset to extract the most useful variables affecting energy consumption. After that, the dataset is partitioned into a set of clusters and be applied a separate prediction model on each cluster.
\vspace{-0.5cm}
\begin{figure}[H]
    \centering
    \includegraphics[width=1\linewidth]{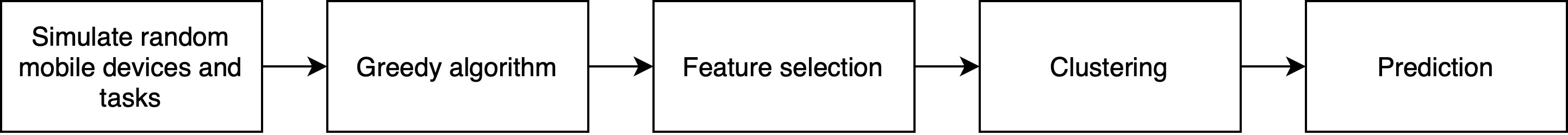}
    \caption{Pipeline structure of the proposed mechanism}
    \label{fig:pipeline}
\vspace{-1 cm}
\end{figure}

\vspace{-0.2 cm}\subsubsection{Feature selection process}\vspace{-0.21 cm}
\label{sec:DesignAlgClusterPhase}
Due to the non-linear relationships between the other features and energy consumption, the correlation metrics fail to detect the effect of these variables on the energy. We employs the algorithm to calculate the MI between variables and energy named \textit{mutual\_information\_regression} \cite{scikit-learn}.

\begin{figure}[!h]
    \centering
\begin{subfigure}{.5\textwidth}
  \centering
  \includegraphics[width=\linewidth,trim={0 1.5cm 0 2.5cm},clip]{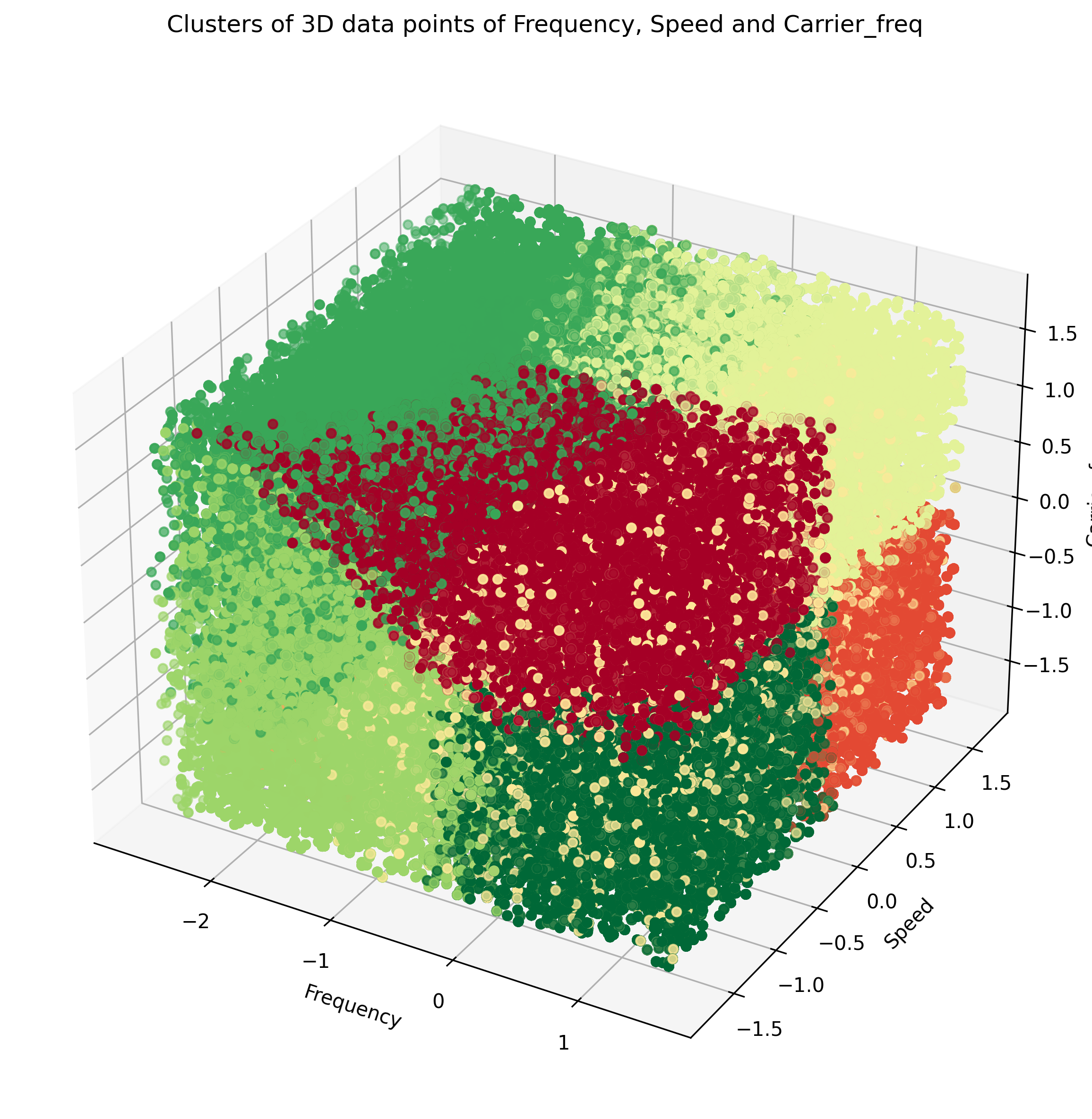}
  \vspace{-0.3 cm}
  \caption{3D clustering structure}
  \label{fig:PremClusteringExamsub1}
\end{subfigure}%
\begin{subfigure}{.5\textwidth}
  \centering
  \includegraphics[width=0.9\linewidth,trim={0.5cm 0.9cm 0.9cm 0.5cm},clip]{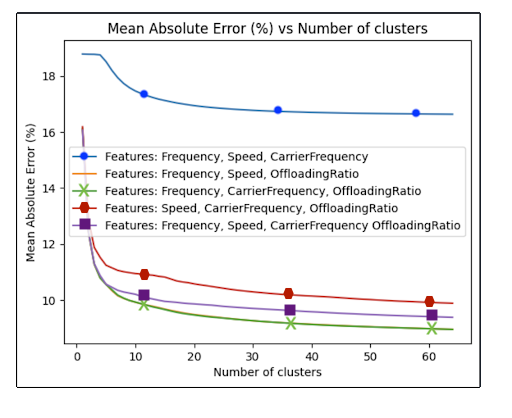}
  \vspace{-0.3 cm}
  \caption{MAE clustering examination}
  \label{fig:PremClusteringExamsub2}
\end{subfigure}
\vspace{-0.5 cm}
    \caption{Clustering examination}
    \label{fig:PremClusteringExam}
\vspace{-0.9 cm}
\end{figure}

In addition, Figure \ref{fig:PremClusteringExamsub1} indicates the possibility of clustering to help in prediction model. This figure demonstrates how data points are grouped into clusters, which can be essential in identifying patterns and is a more intuitive grasp of the data distribution. The use of clustering can simplify the data pre-processing steps, making it easier to handle large datasets and improve the overall efficiency of the development process with the selected features as shown in Figure \ref{fig:PremClusteringExamsub2}.

\begin{remark} \textbf{(Clustering MI Impact)}
\label{theory:remarkStrongClusterImpact}
\setstretch{1}
We further investigate the concern on feature clustering on our greedy result which is shown in Figure \ref{fig:PremClusteringExam}. This preliminary result is important in illustrating that extracting all 4 features does not achieve the best MAE error rate. In contrast, extracting a subset of 3 out of 4 features might be less accurate than using a larger number of features. However, when an appropriate subset of features is selected, where the clustering ability is stronger, indicated by a higher MI value, the final error rate can be even better than when using all 4 features.
\end{remark}

After pre-traning step, it is common to make an assumption of the expected output monotonicity. At each step, the given solution is the most valuable subset of items that satisfy the constraints. Now if the capacity increases, the new set of item about to be selected are expected to be at least as valuable as before
\begin{assumption} 
\label{ass:monosssumpt}
\setstretch{1}
\textbf{(Theorem 1 in \cite{gergatsouli2020black})}
In a setting of d dimension feature, there exists a meta-algotithm $\mathcal{M}_q$ that corrects the monotonicity for any function K-mean cluster c: $\mathbb{R}^d \rightarrow [0,1]^K$ has $E[\mathcal{M}_c]\geq \mathbb{E}[q(c)]-\epsilon$, where the first expectation is taken over the input distribution and the randomness in the meta-algorithm
\end{assumption}

\begin{assumption} 
\label{ass:pretrainCond}
\setstretch{1}
At each state of the training process iterates over a set state $S_t, 0 \leq t \leq T$, the context environment is $e_{i}$ and cluster algorithm propose a result of $c \sim q(c)$ clustering distribution and there is existed a limit $\beta$: $logP(c) \leq \beta$ which a normal situation to ensure the effectiveness of clustering pre-training.
\end{assumption}

The concurrent probability of context environment over set of cluster
\vspace{-0.2cm}
\begin{equation}
    \label{eq:concContexOverCluster}
    \mathbb{P}(c|S_t)
    \overset{(a1)}{=}\frac{P(S_t,c)|P(c)}{\sum\limits_{c' \in q(c)} P(S_t,c')P(c')dc'}\overset{(a2)}{=}\frac{\prod_{i=0}^{t}P(r_i|c,S_t,e_i)P(c)}{\sum\limits_{c' \in q(c')} \prod_{i=0}^{t}P(r_i|c',S_t,e_i) P(c')dc'}
\end{equation}
\vspace{-0.5cm}

The pre-training with 2 condition (a1) span over clustering $c \in q(c)$ at each stage $S_t$ and (a2) the concurrent probability of context environment span over set of cluster $e_i$

\vspace{-0.8cm}

\begin{proposition} \textbf{The error bound of clustering prediction}
\label{theory:ClusteringProposition}
\setstretch{1.1}
The boundness of the probability of reward $r_{t+1}$
\vspace{-0.3cm}
\begin{equation}
\label{propBound}
\mathbb{E}_{Q}\mathbb{P}(r_{i}|\cdot,e_{i+1}))-\mathbb{E}_{Q} \mathbb{P}(r_i|\cdot,e_i) \leq \beta    
\vspace{-0.3cm}\end{equation}
where it employs the bound of $P(c)$, i.e. $logP(c) \leq \beta$
\end{proposition}

\vspace{-0.5 cm}
\renewcommand*{\proofname}{Sketch of proof}
\begin{proof}
\setstretch{1.1}
We roll out the probability of the reward in (\ref{eq:reward})
\vspace{-0.2 cm}
\begin{equation}
\label{eq:reward}
-\sum_{t=0}^T\log(\mathbb{P}(r_{t+1}|S_t,e_{i}))
\vspace{-0.5cm}\end{equation}

\noindent
using where (b1) is Bayesian model averaging, (b2) is the substitution of (\ref{eq:concContexOverCluster})
and the properties of integral $\int_{c \in q(c)}F_X(c)q(c)dc=\mathbb{E}_{c\sim q(c)}F_X(c)$ and we have the monotonicity from the common assumption \ref{ass:monosssumpt} of meta-learning (MAML) then we have $Inf_{P} F(f,P(S)) = \int_S f(s,P(S))\mu(S)dS$ as an appliance of Riemann integral definition. We arrive at

\vspace{-0.7 cm}
\begin{equation}
\label{eq:tempRepPro}
    -\sum_{t=0}^T\log(\mathbb{P}(r_{t+1}|S_t,e_{i}))\geq -\mathbb{E}_{c\sim q(c)} \mathbb{P}(r_i|c,S_t,e_i) - \mathbb{E}_{c\sim q(c)}logP(c)
\end{equation}
\vspace{-0.5 cm}

Then we rewrite (\ref{eq:tempRepPro}) in average form to obtain (\ref{propBound}). Further more, we rewrite in the KL as usual formulation in K-cluster domain.

\vspace{-0.5 cm}
{\allowdisplaybreaks
\begin{align*}
\label{eq:tempLemma2}
\mathbb{E}_{c\sim q(c)} \left[ KL(\mathbb{P}(r_{t+1}|\cdot,e_{i+1}) \| \mathbb{P} (r_i|\cdot,e_i))\right] &\leq \beta
\end{align*}
}
\vspace{-0.5 cm}
\end{proof}

\vspace{-1 cm}
\begin{algorithm}[!h]
    \caption{Algorithm \textbf{ClusPreAlg} based on Clustering-Prediction technique} 
    \label{alg:cluster_predict}
    \small
    \begin{algorithmic}[1]
%
        \Function{TrainClusteredModels}{num\_clusters}
            \State Apply K-means clustering to segment the data into \textit{num\_clusters} clusters
            \State Train a prediction model for each cluster
            \State \Return \{models\_per\_cluster, cluster\_classifier\}
        \EndFunction

        
        \Function{ClusteringPredict}{models, classifier, test\_data}
            \State Determine the cluster for \textit{test\_data} using \textit{classifier}
            \State Retrieve the prediction model for the identified cluster
            \State Use the model to predict the energy value for \textit{test\_data}
            \State \Return predicted\_energy\_value
        \EndFunction

        
        \Function{EvaluateModels}{training\_data, test\_data, k}
            \For{\textit{num\_clusters} from 1 to \textit{k}}
                \State \{models, classifier\} $\gets$ \Call{TrainClusteredModels}{num\_clusters}
                \State predictions $\gets$ \Call{ClusteringPredict}{models, classifier, test\_data}
            \EndFor
            
        \EndFunction
    \end{algorithmic} 
\end{algorithm}
\vspace{-0.2 cm}
\linespread{1.4}

\vspace{-0.2 cm}\subsubsection{The clustering prediction ClusPreAlg algorithm}\vspace{-0.21 cm}
\label{sec:DesignAlgClusterPredict}
Equation (\ref{formula:e_total}) shows no optimum solution, we devise a numerical approximation method to predict the total energy consumption. The model has three steps:
\linespread{1.12}
\vspace{-0.2 cm}
\begin{itemize}
    \item Scaling the variables to the range [0,1]
    \item Performing the K-means clustering with given number of clusters $k$
    \item Train separate prediction models on each cluster
\vspace{-0.2 cm}
\end{itemize}

Different variables have distinct units and interpretations. For example, a speed of $1 m/s$ is not equivalent to a processing frequency of $1 Hz$. Scaling transform eliminates concerns about their actual units. Additionally, K-means clustering minimizes the inertia function, which is heavily influenced by the scale of the features. 
As a result, features with strong clustering can dominate and overshadow those with smaller ones.
The algorithm is given as follows:

\vspace{-0.2 cm}
\section{Numerical results}
\vspace{-0.4 cm}\subsection{Experiment set-up}
\begin{figure}[ht]
    \centering
    \includegraphics[width=0.4\linewidth]{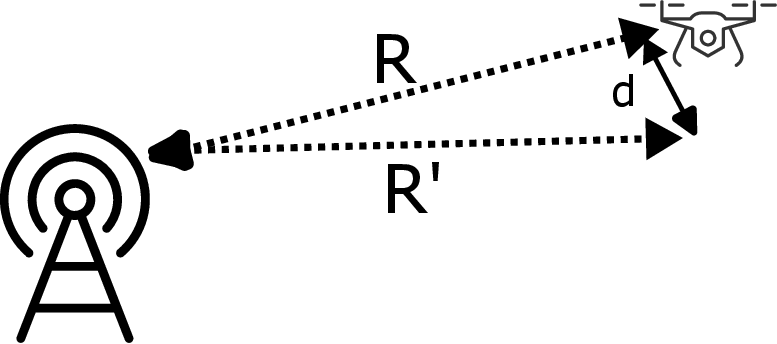}
    \vspace{-0.3cm}
    \caption{The moving distant vs signal radius distant}
    \label{fig:traRep}
\vspace{-0.7cm}
\end{figure}

\vspace{-0.21 cm}
We leverage a novel large-scale database of a tracking data set in a real vehicle trajectory. The VED dataset \cite{ved}, which logs GPS trajectories of automobile in timing recording and monitoring the status of fuel, energy, speed and auxiliary power consumption.
Since our clustering prediction primarily concern the value of mobile/IoT device traveling in 2D earth coordinator of longitude and latitude. 
To avoid the partial interference derivation between the two dimension in velocity estimation, we are interested in discovering the feature of device modeling as UAV in plane where the moving distance $d << R $ as shown in Figure \ref{fig:traRep}, then we can transform that $d \approx R \times \phi$ We develop a Python library named Spectral Efficiency Estimator as in section \ref{sec:sysArch}, which is deployed on a server run at 3.6GHz, 16 cores and 32 GB memory. The dataset VED is spanned on 12GB size which includes 22.3 million of records and be accessed through Python Pandas library \cite{reback2020pandas,harris2020array}.

\vspace{-0.7cm}

\vspace{-0.4 cm}\subsection{The evaluation of the baseline IterGAlg algorithm}\vspace{-0.21 cm}

\subsubsection{The impact of modulation settings}
The system's efficiency is measured by the total energy consumption of MDs and processing of their tasks. The size of data need to be uploaded has an important impact on the overall system performance. To evaluation, we derive the following experiment to figure out the data size impact.

\begin{figure}[H]
    \centering
    \includegraphics[width=0.7\linewidth,,height=5cm]{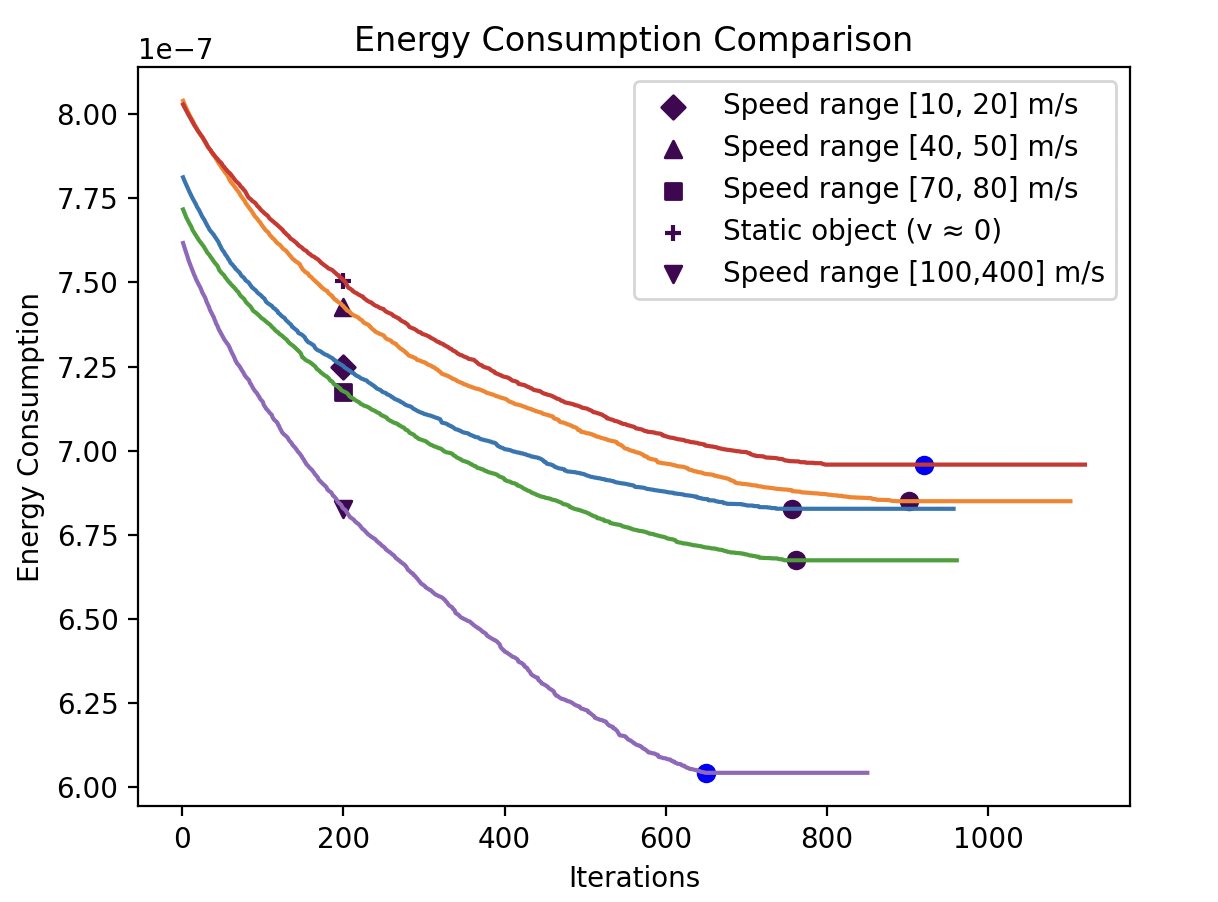}
\vspace{-0.3 cm}
    \caption{Total energy consumption in various modulation setting}
    \label{fig:exp-greedy}
\end{figure}

The proposed algorithm is validated in Figure. \ref{fig:exp-greedy} here ensures for local optimization and may not guarantee a global minimum due to their inherent approach of selecting the best immediate option. It also shows that in high mobility scenario, as speed ranging from 100 - 400 m/s and the spectral efficiency drops, results in lower energy consumption overall. This could be explained due to the reduced communication rate without power adjust to balance the rate loss.

\vspace{-0.8 cm}

\vspace{-0.8 cm}
\subsubsection{The impact of the offloading ratio}
\begin{figure}[H]
\centering
\includegraphics[width=0.5\textwidth]{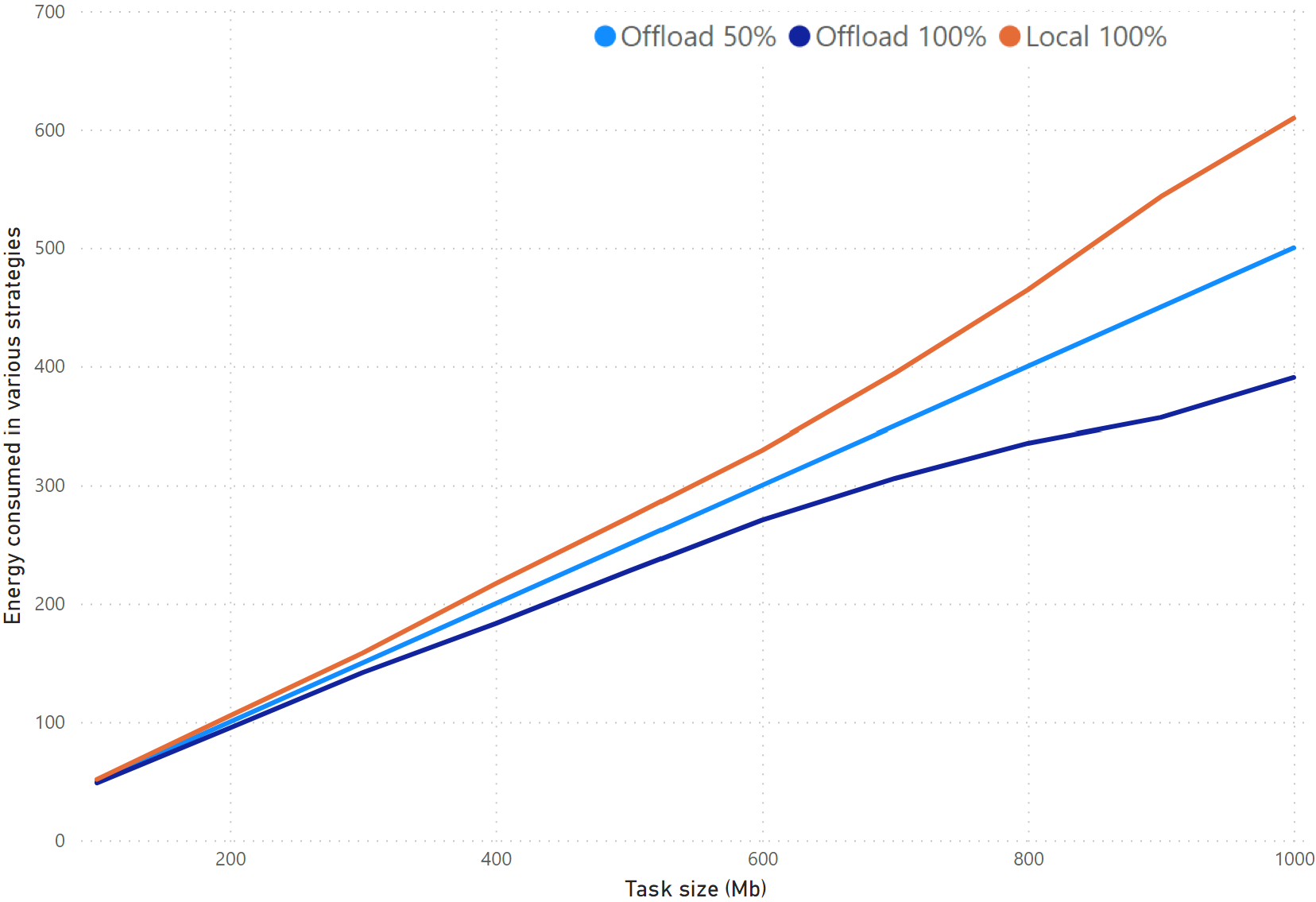}
\vspace{-0.2 cm}
\caption{Impact of data size on the total energy consumption} 
\label{fig:DataSizeEnergyConsumption}
\vspace{-0.2cm}
\end{figure}

The greedy algorithm aims to reduce the total energy by taking an iterative approach to an optimal solution as it selects locally optimal energy. As seen in the previous evaluation, the greedy offloading method makes choices based on immediate reduction, which does surely lead to a convergence outcome. Besides, the impact of data size on energy consumption must be taken into account since it is crucial to the edge data-driven network. The energy needed for data transmission grows significantly as the size of the data increases exponentially. 
Since mobile edge computing typically prefers to offload tasks to server, it is mandatory to determine how to reduce energy usage while still achieving optimal results in a limited amount of time. 
The better energy saving gap in increasing the offloading data portion is confirmed by the improved gap (the larger gap is the better) as in Figure \ref{fig:DataSizeEnergyConsumption}.

\vspace{-0.4 cm}

\begin{figure}[!h]
    \centering
\begin{subfigure}{.5\textwidth}
  \centering
  \includegraphics[width=\linewidth]{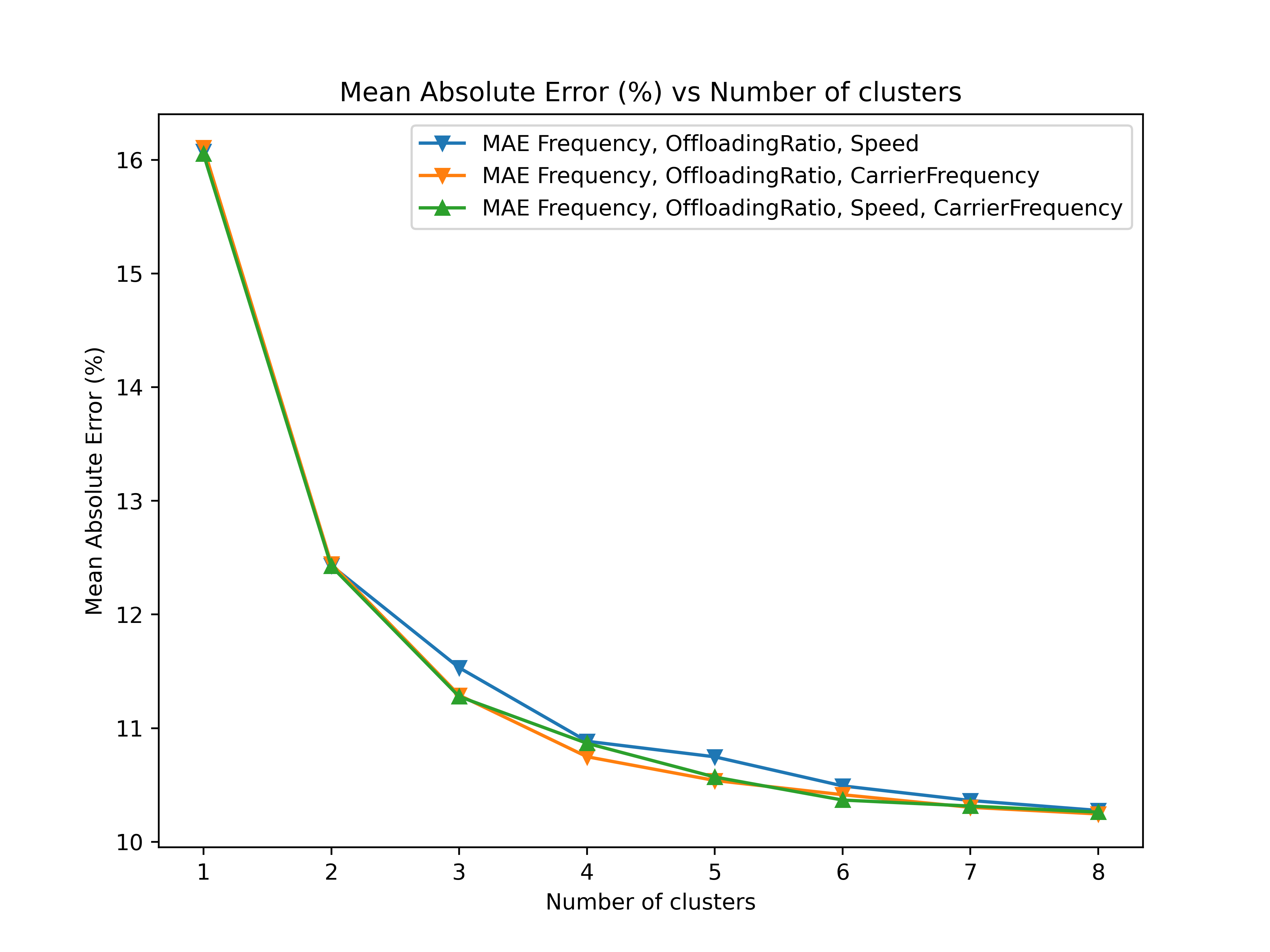}
  \vspace{-0,8 cm}
  \caption{MAE}
  \label{fig:MAEsub1}
\end{subfigure}%
\begin{subfigure}{.5\textwidth}
  \centering
  \includegraphics[width=\linewidth]{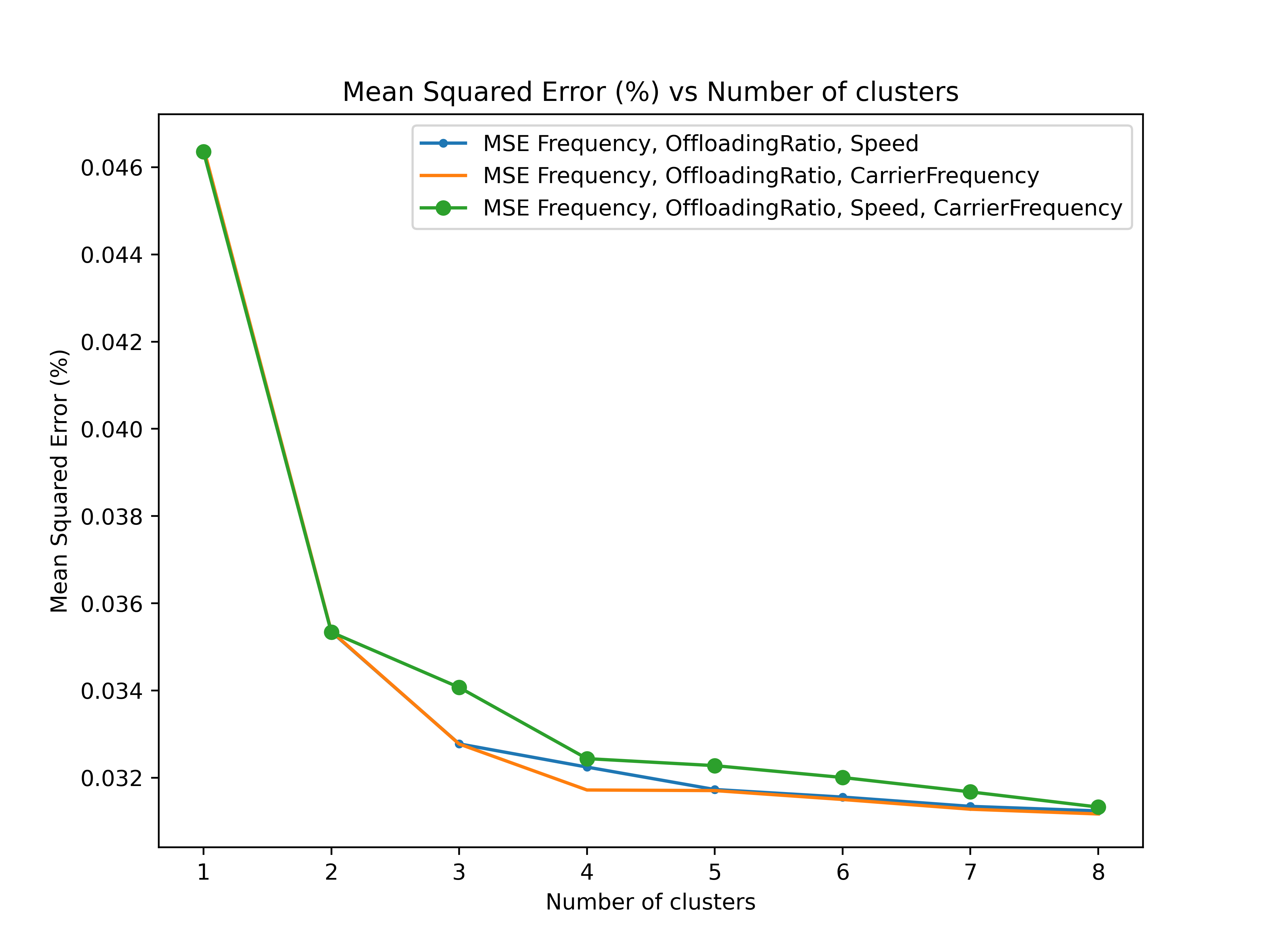}
  \vspace{-0.8 cm}
  \caption{MSE}
  \label{fig:MSEsub2}
\end{subfigure}
\vspace{-0.5 cm}
    \caption{MAE and MSE with different $k$ values}
    \label{fig:MAE_result}
\end{figure}

\vspace{-0.4 cm}
\subsection{The evaluation of the enhanced ClusPreAlg algorithm}\vspace{-0.21 cm}
In this assessment, we use the greedy numerical outcomes as a baseline to evaluate the enhanced clustering prediction. We aim to assess the error rate in term of MAE and MSE for the predicted results and compare them with the baseline numerical data.

Figure \ref{fig:MAEsub1} illustrates that the Mean Absolute Error (MAE) shows a substantial reduction by 10\%, highlighting the algorithm's improved accuracy.
\noindent
The effects of \textit{Carrier Frequency} and \textit{Speed} on energy consumption are consistent with the observations in \ref{sec:DesignAlgClusterPhase}. Consequently, the curve of all features does not change significantly when less influential factors such as are added.
Figure \ref{fig:MSEsub2} emphasizes the same pattern observed with MAE but with significantly better results, as the Mean Squared Error (MSE) is as low as 3\%. This highlights the effectiveness of the clustering approach in minimizing prediction errors more robustly when considering the squared differences.

\begin{figure}[h]
    \centering
    \includegraphics[width=0.5\linewidth]{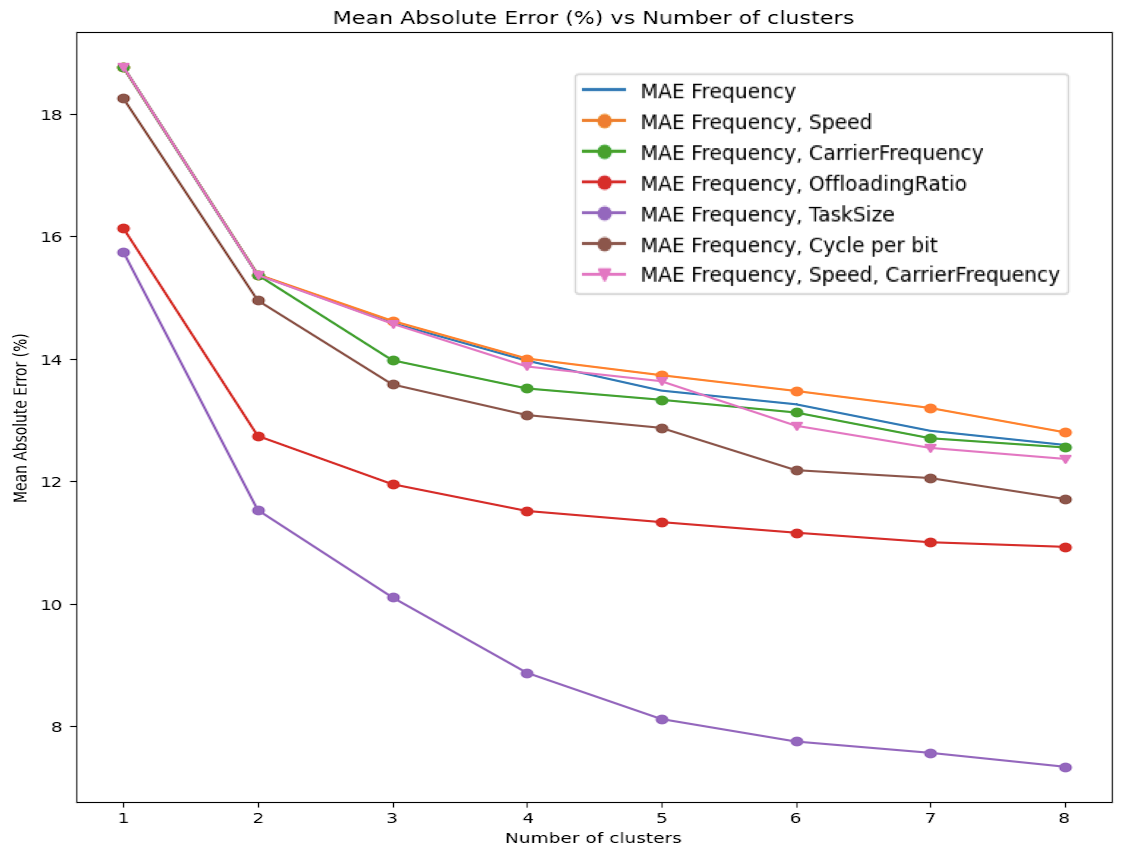}
\vspace{-0.3 cm}
    \caption{MAE evaluation of full feature clustering}
    \label{fig:fullFeatureMAE}
\end{figure}

To confirm the effect of featured clustering, we perform a full feature clustering in Figure \ref{fig:fullFeatureMAE}. This confirm the most impact factor of the mobile task includes TaskSize and OffloadingRatio. 
Therefore, our proposed mechanism has an impact on resource allocation, particularly in IoT systems with heavy computational task and require fast processing step by leveraging prediction techniques.

Through historical optimized offloading decisions kept in a dataset, a mutual information analysis confirmed the relationship between the variables and energy consumption. Subsequently, k-means clustering applied to the dataset revealed distinct clustering characteristics, enabling the construction of separate prediction models for each cluster. These models achieved low level errors in terms of MAE and MSE. As a result, the clustering - prediction model could serve as an improvement over the baseline greedy algorithm while requiring much less variable information.

\section{Conclusion}
\vspace{-0.2cm}
In IoT system deployment, the computation offloading leverages the powerful capabilities of nearby computation of the base station at the edge network.
This study reveals significant impacts of offloading decisions on total energy consumption in MEC network employing OTFS modulation. We propose a number of comprehensive models for mobile devices, tasks, local energy computation, and offloading energy computation. From these models, optimized offloading strategies by the greedy algorithm demonstrate a clear potential for reducing energy use, confirmed by convergence plots showing minimized energy. 

We analyze the fundamentals of offloading within the new velocity modulation scheme, which figures out the OTFS network according to time, frequency, and space. The approximation through a clustering - prediction model  show the potential of adaptive mechanism in resource allocation for ISAC enabled system where sensing and communication have strongly interaction.
However, we lack a definitive strategy for choosing the starting point of the greedy algorithm, which affects our ability to ensure a good initial point and establish a bounded statement of convergence.

\bibliographystyle{./styles/bibtex/splncs03_unsrt}
\renewcommand{\bibname}{References}
\bibliography{author}

\begin{thebibliography}{10}
\providecommand{\url}[1]{\texttt{#1}}
\providecommand{\urlprefix}{URL }

\bibitem{xiong2023fundamental}
Xiong, Y., Liu, F., Cui, Y., Yuan, W., Han, T.X., Caire, G.: On the fundamental tradeoff of integrated sensing and communications under gaussian channels. IEEE Transactions on Information Theory  (2023)

\bibitem{kovachev2012adaptive}
Kovachev, D., Yu, T., Klamma, R.: Adaptive computation offloading from mobile devices into the cloud. In: 2012 IEEE 10th International Symposium on Parallel and Distributed Processing with Applications. pp. 784--791. IEEE (2012)

\bibitem{ma2017efficient}
Ma, L., Yi, S., Li, Q.: Efficient service handoff across edge servers via docker container migration. In: Proceedings of the Second ACM/IEEE Symposium on Edge Computing. pp. 1--13 (2017)

\bibitem{mach2017mobile}
Mach, P., Becvar, Z.: Mobile edge computing: A survey on architecture and computation offloading. IEEE communications surveys \& tutorials  19(3),  1628--1656 (2017)

\bibitem{nguyen2019computation}
Nguyen, P., Ha, V., Le, L.: Computation offloading and resource allocation for backhaul limited cooperative mec systems. In: Proc. IEEE VTC Fall. Hawaii, USA (Sep 2019)

\bibitem{nguyen2020joint}
Nguyen, P.D., Le, L.B.: Joint computation offloading, sfc placement, and resource allocation for multi-site mec systems. In: 2020 IEEE Wireless Communications and Networking Conference (WCNC). pp. 1--6. IEEE (2020)

\bibitem{hadani2017orthogonal}
Hadani, R., Rakib, S., Tsatsanis, M., Monk, A., Goldsmith, A.J., Molisch, A.F., Calderbank, R.: Orthogonal time frequency space modulation. In: 2017 IEEE Wireless Communications and Networking Conference (WCNC). pp. 1--6. IEEE (2017)

\bibitem{raviteja2019orthogonal}
Raviteja, P., Phan, K.T., Hong, Y., Viterbo, E.: Orthogonal time frequency space (otfs) modulation based radar system. In: 2019 IEEE Radar Conference (RadarConf). pp. 1--6. IEEE (2019)

\bibitem{rajeswaran2019meta}
Rajeswaran, A., Finn, C., Kakade, S.M., Levine, S.: Meta-learning with implicit gradients. Advances in neural information processing systems  32 (2019)

\bibitem{zak_transform}
Mohammed, S.K.: Time-domain to delay-doppler domain conversion of otfs signals in very high mobility scenarios. IEEE Transactions on Vehicular Technology  70(6),  6178--6183 (2021)

\bibitem{nguyen2012ebc}
Nguyen, D., Thoai, N.: Ebc: Application-level migration on multi-site cloud. In: 2012 International Conference on Systems and Informatics (ICSAI2012). pp. 876--880. IEEE (2012)

\bibitem{chen2019dynamic}
Chen, M., Li, W., Fortino, G., Hao, Y., Hu, L., Humar, I.: A dynamic service migration mechanism in edge cognitive computing. ACM Transactions on Internet Technology (TOIT)  19(2),  1--15 (2019)

\bibitem{sonkoly2020scalable}
Sonkoly, B., Haja, D., N{\'e}meth, B., Szalay, M., Czentye, J., Szab{\'o}, R., Ullah, R., Kim, B.S., Toka, L.: Scalable edge cloud platforms for iot services. Journal of Network and Computer Applications  170,  102785 (2020)

\bibitem{suganuma2018multiagent}
Suganuma, T., Oide, T., Kitagami, S., Sugawara, K., Shiratori, N.: Multiagent-based flexible edge computing architecture for iot. IEEE Network  32(1),  16--23 (2018)

\bibitem{scikit-learn}
Pedregosa, F., Varoquaux, G., Gramfort, A., Michel, V., Thirion, B., Grisel, O., Blondel, M., Prettenhofer, P., Weiss, R., Dubourg, V., Vanderplas, J., Passos, A., Cournapeau, D., Brucher, M., Perrot, M., Duchesnay, E.: Scikit-learn: Machine learning in {P}ython. Journal of Machine Learning Research  12,  2825--2830 (2011)

\bibitem{gergatsouli2020black}
Gergatsouli, E., Lucier, B., Tzamos, C.: Black-box methods for restoring monotonicity. In: International Conference on Machine Learning. pp. 3463--3473. PMLR (2020)

\bibitem{ved}
Oh, G., Leblanc, D.J., Peng, H.: Vehicle energy dataset (ved), a large-scale dataset for vehicle energy consumption research. IEEE Transactions on Intelligent Transportation Systems  23(4),  3302--3312 (2020)

\bibitem{reback2020pandas}
Reback, J., McKinney, W., Van Den~Bossche, J., Augspurger, T., Cloud, P., Klein, A., Hawkins, S., Roeschke, M., Tratner, J., She, C., et~al.: pandas-dev/pandas: Pandas 1.0. 5. Zenodo  (2020)

\bibitem{harris2020array}
Harris, C.R., Millman, K.J., Van Der~Walt, S.J., Gommers, R., Virtanen, P., Cournapeau, D., Wieser, E., Taylor, J., Berg, S., Smith, N.J., et~al.: Array programming with numpy. Nature  585(7825),  357--362 (2020)

\end{thebibliography}
\end{document}